\documentclass[%
 reprint,
 amsmath,amssymb,
 aps,
prl
]{revtex4-1}

\usepackage{graphicx}
\usepackage{dcolumn}
\usepackage{bm}
\usepackage{color}
\usepackage{hyperref}
\hypersetup{
  colorlinks   = true,  
  urlcolor     = blue,  
  linkcolor    = blue,  
  citecolor    = red    
}

\begin{document}

\preprint{APS/123-QED}

\title{Atomic Clock Measurements of Quantum Scattering Phase Shifts \\
Spanning Feshbach Resonances at Ultralow Fields}

\author{Aaron Bennett}
\affiliation{Department of Physics, The Pennsylvania State University, University Park, PA, 16802, USA}
\author{Servaas Kokkelmans}
\affiliation{Eindhoven University of Technology, P. O. Box 513, 5600 MB
Eindhoven, The Netherlands}
\author{Jeremy M. Hutson}
\email{J.M.Hutson@durham.ac.uk} \affiliation{ Joint Quantum Centre (JQC)
Durham-Newcastle, Department of Chemistry, Durham University, South Road,
Durham, DH1 3LE, United Kingdom }
\author{Kurt Gibble}
\email{kgibble@psu.edu}
\affiliation{Department of Physics, The Pennsylvania State University, University Park, PA, 16802, USA}

\date{\today}

\begin{abstract}
We use an atomic fountain clock to measure quantum scattering phase shifts
precisely through a series of narrow, low-field Feshbach resonances at average
collision energies below $1\,\mu$K. Our low spread in collision energy yields
phase variations of order $\pm \pi/2$ for target atoms in several $F,m_F$
states. We compare them to a theoretical model and establish the accuracy of
the measurements and the theoretical uncertainties from the fitted potential.
We find overall excellent agreement, with small statistically significant
differences that remain unexplained.



\end{abstract}

\maketitle


Coherence and the precise measurements allowed by long coherence times are
central themes in atomic physics. The coherent nature of atom-atom scattering
is important in phenomena such as Bose-Einstein condensation \cite{Hall98,
Widera06}, Feshbach resonances \cite{Chin10, Inouye98, Chin04} and ultracold
molecule formation \cite{Regal:40K2:2003, Herbig:2003}. Atom-atom scattering
also shifts the frequency of atomic clocks and interferometers \cite{Gibble93,
Fertig00, PeirerraDosSantos, Szymaniec07, Papoular12}, which often limits their
precision and accuracy. Conversely, atom interferometry can directly probe the
phase shifts at the core of quantum scattering
\cite{Legere98,Wilson04,Walraven04,Rempe05,Wilson07,Hart07,Gensemer12} and
sensitively test models of atom-atom interactions.

Accurate knowledge of low-energy scattering is especially important for cesium
as its clock collisional frequency shift is predicted to pass through zero
around 100 nK \cite{Szymaniec07}.  This is the energy scale for collisions in
PHARAO, a microgravity laser-cooled cesium clock scheduled to launch soon as
part of the ACES mission \cite{Laurent06}. Additionally, precise measurements
of scattering phase shifts, or equivalently scattering lengths, near narrow
Feshbach resonances may provide high sensitivity to the time
variation of fundamental constants \cite{Chin06, Borschevsky11}.

Here we use an atomic clock to make precision measurements of phase shifts for
the scattering of ultracold cesium atoms through several narrow Feshbach
resonances, as the magnetic field increases from 0 to 0.4 G. A narrow spread of
collision energies allows us to observe phase shift variations of nearly $\pi$
through the resonances. We establish the accuracy of our measurements and
compare them to coupled-channel calculations that use recent interaction
potentials from fits to Feshbach resonances and near-threshold bound states at
fields from $10$~G to $1000$~G \cite{Berninger:Cs2:2013}. We find overall
excellent agreement with the model for the positions of the ultra-low-field
Feshbach resonances, significantly improved from that obtained using the
previous best interaction potential \cite{Chin04}. The absolute phase-shift
differences also agree well, although some scattering channels show significant
and yet-unexplained deviations.

\begin{figure*}[t]
  \centerline{\includegraphics[scale=.64]{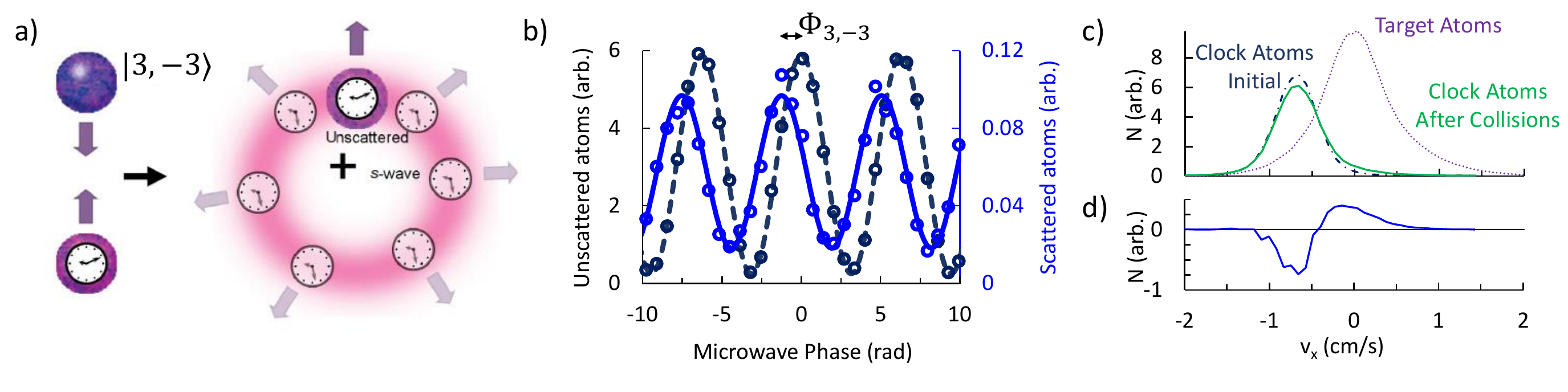}}
\caption[Probe Velocities for Interference Current Check.]{ (a) Atoms in a
coherent superposition of the cesium clock states collide with atoms in a
target state. The clock atom wave packet scatters as a spherically outgoing
s-wave (pink shell) and continues unscattered (violet cloud). The
clock faces indicate the differential scattering phase shift of the clock
coherence. (b) Transition probability for scattered (solid blue) and
unscattered (dashed grey) clock atoms. Each data point represents a single
fountain launch with target atoms in $\vert 3, -3 \rangle$, a mean collision
energy of $E_c=798$ nK, and $B=80$ mG. (c) After the first Ramsey pulse, the
clock atoms prepared with dash-dot blue velocity distribution collide with the
target atoms with the dotted purple distribution. The collisions redistribute
the clock atom velocities (solid green). d) Subtracting the initial clock
velocity distribution from the distribution with scattering shows the net
redistribution, which shifts the initial velocity class towards $v=0$. The mean
collision energy can be tuned by changing the initial selected and final
detected velocity.} \label{fig:fig1}
\end{figure*}

Our interferometric technique \cite{Hart07,Gensemer12} precisely
and unambiguously detects differences of quantum scattering phase shifts
\cite{Legere98,Wilson04, Walraven04,Rempe05,Wilson07}. Such
information is difficult to extract from measurements of scattering cross
sections, both because cold atom densities are challenging to measure
accurately and because cross sections depend on the squares of scattering
lengths. In our atomic fountain clock, a microwave $\pi$/2 pulse creates a
coherent superposition of the cesium clock states $\vert F, m_F \rangle = \vert
3, 0\rangle \equiv 3$ and $\vert 4, 0\rangle \equiv 4$. The clock atoms then
collide with ``target'' atoms in another state $\vert j \rangle \equiv \vert F,
m_F \rangle$ with s-wave phase shifts $\delta_{3,j}$ and $\delta_{4,j}$,
forming an outgoing spherical shell as shown in Fig.~\ref{fig:fig1}a.
Consequently, the phase of the scattered clock coherence, represented by the
clock hands in Fig.~\ref{fig:fig1}a, jumps by the difference of the s-wave
phase shifts, $\Phi_j=\delta_{4,j}-\delta_{3,j}$. A second $\pi$/2 pulse with
an adjustable phase yields a Ramsey fringe with the phase shift of the clock
coherence $\Phi_j$. The scattered atoms are detected, and the atoms in the
forward scattering direction excluded, using a velocity-selective stimulated
Raman transition \cite{Kasevich91}. This technique takes advantage of the phase
and frequency accuracy of atomic clocks and precisely probes arbitrarily large
phase differences. Several other techniques have also been demonstrated that
precisely probe small differences of scattering lengths \cite{Widera06,
Matthews98, Egorov13}.

Scattering phase shifts change by $\pi$ as the magnetic field $B$ is scanned
across a Feshbach resonance. However, observing the full phase variation
requires a narrow spread of collision energies. Our previous observations of
cesium scattering phase shifts through Feshbach resonances studied the
scattering between atoms in two distinct clouds
\cite{Gensemer12} in our juggling atomic clock \cite{Legere98}. At collision
energies $E_{\rm c}$ between 12 and 50 $\mu$K, cloud temperatures even as low
as 400 nK give a significant spread of collision energies, of order 10 $\mu$K,
broadening the narrow resonances and suppressing the excursions of the phase
shifts \cite{Gensemer12}. Here, we instead select and collide two
velocity classes from a single launched cloud in our fountain clock. The low
collision energies of 0.5 to 1 $\mu$K and correspondingly narrow energy spread
yield observed phase-shift variations of nearly $\pi$ through several narrow
Feshbach resonances.

Our experimental sequence begins with launching atoms from a magneto-optical
trap and cooling them to 400 nK with degenerate sideband cooling in a
moving-frame 3D optical lattice
\cite{Hart07,Gensemer12,Treutlein01}. After the sideband lattice
cooling, 65\% of the atoms are in $\vert 3, 3\rangle$, 20\% are in $\vert 3, 2
\rangle$, and the rest are in other $\vert 3, m_F \rangle$ states. The atoms in
$\vert 3, 3 \rangle$ are transferred to the desired target state, $\vert 3,
m_F=\pm1, \pm 2, \pm 3 \rangle$ or $\vert 4, m_F\neq 0 \rangle$, by a series of
microwave pulses. To prepare $m_F<0$ target states, a non-adiabatic magnetic
field reversal precedes the microwave pulses to transfer the atoms from $\vert
3, 3\rangle$ to $\vert 3, -3\rangle$. For all targets except $\vert 3, \pm
1\rangle$ \cite{ABThesis} \footnote{See Supplemental Material for
experimental details.}, the atoms initially in $\vert 3, 2 \rangle$ are
transferred to either $\vert 3, 0 \rangle$ or $\vert 4, 0 \rangle$ by another
series of microwave pulses, interleaved with the target-atom microwave pulses,
and a stimulated Raman pulse. The Raman pulse is velocity-sensitive and selects
a slice of the velocity distribution, 36 nK wide, in the
horizontal $x$ direction, imparting two photon recoils to the selected atoms,
as in Fig.~\ref{fig:fig1}c. Unwanted atoms in other $m_F<0$ states and other
velocity classes are removed with clearing laser pulses tuned to the
6$S_{1/2} \rightarrow 6P_{3/2}$, $F=3\rightarrow 5'$ and $3
\rightarrow 2'$ transitions. A $\pi/2$ microwave pulse then prepares the clock
atoms in a coherent superposition of $\vert 3, 0 \rangle$ and $\vert 4, 0
\rangle$, after which the collisions of the clock atoms with the target atoms
above the clock cavity change their velocities $v$. In Fig.~\ref{fig:fig1}c,
the collisions tend to scatter atoms with large velocities towards $v=0$
\cite{Gibble95} as they begin to thermalize. For the small fraction of clock
atoms that scatter, the phase of the clock coherence is shifted by the
difference of the s-wave scattering phase shifts \cite{Hart07}.
After the atoms fall back into the cavity, a second microwave $\pi/2$ pulse
produces the Ramsey fringe in Fig.~\ref{fig:fig1}b. A clearing pulse removes
the target atoms, as well as the clock atom population in the same hyperfine
state $F$ as the target atoms. For $\vert 4, m_F \rangle$ target atoms, a
stimulated Raman transition (the Raman probe) transfers a narrow,
36 nK wide, velocity class of scattered atoms to $\vert 4, 0
\rangle$. A laser resonant with the 4 $\rightarrow 5'$ transition excites these
atoms and we collect their fluorescence to obtain Fig.~\ref{fig:fig1}b. In
Fig.~\ref{fig:fig1}b, we also measure a reference Ramsey fringe, where we clear
the target atoms before the first Ramsey pulse and detect atoms at the center
of the clock-atom velocity distribution. For $\vert 3, m_F \rangle$ target
atoms, an additional microwave pulse after the $F=3$ clearing pulse transfers
the clock atoms in $\vert 4, 0 \rangle$ to $\vert 3, 0 \rangle$, and then a
second clearing pulse removes $F=4$ atoms before a stimulated Raman probe as
above. We evaluate and subtract backgrounds using a pump-probe technique that
clears the target atoms immediately before the first Ramsey pulse, inhibits the
clock atom Raman selection, or both, to yield the Ramsey fringes as shown in
Fig.~\ref{fig:fig1}b \cite{Gibble95, Hart07, Gensemer12}.

\begin{figure*}[t]
  \centerline{\includegraphics[scale=.65]{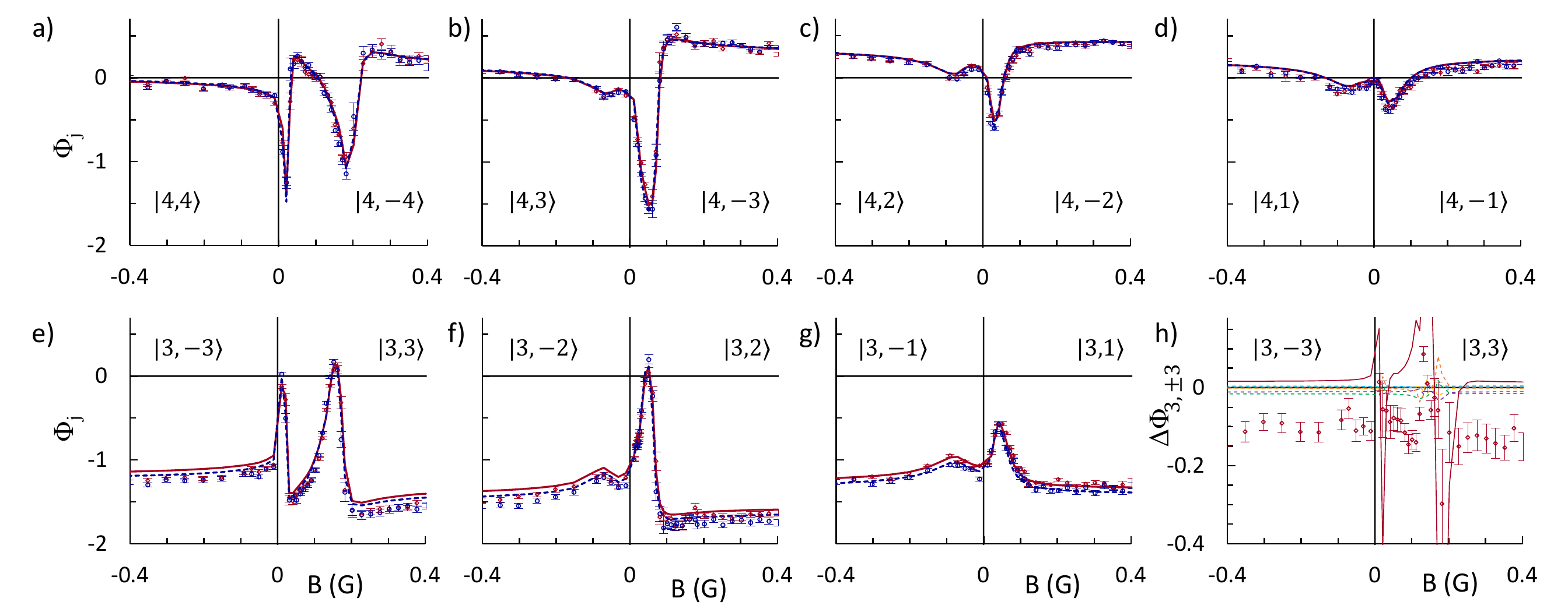}}
\caption[Probe Velocities for Interference Current Check.]{ (color online).
a)-g) Magnetic field dependence of the differential phase shift $\Phi$ for
target atoms in the $\vert 4, m_F\neq 0 \rangle$ and $\vert 3, m_F\neq 0
\rangle$ states; negative $B$ corresponds to the opposite sign of $m_F$.
The scattering phase shifts vary rapidly with magnetic field through
a series of Feshbach resonances. The blue circles (red diamonds) are
experimental results for mean collision energies of 616-656 nK
(746-798 nK) and the curves are corresponding results from
coupled-channel calculations on the best-fit potential
\cite{Berninger:Cs2:2013}.  h) Comparison of measured (red diamonds) and
theoretical values of $\Phi_{3, \pm 3 }$. All are shown after
subtracting $\Phi_{3, \pm 3 }$ from the best-fit potential
\cite{Berninger:Cs2:2013}.  The best-fit potential and
experimental results differ by $\approx 0.1$ rad throughout the
range and their variations through the Feshbach resonances agree
very well. The previous best potential (solid red line) \cite{Chin04} gives
much larger deviations from experiment through the resonances.
The 6 dashed curves indicate the uncertainty of the best-fit potential. Their
differences from the best-fit potential are small compared to
the $\approx 0.1$ rad offset of the experimental results.} \label{fig:fig2}
\end{figure*}

Fig.~\ref{fig:fig2} shows the measured phase shifts for target atoms in each
$\vert F, m_F\neq$ 0 $\rangle$ state, as we traverse a number of low-field
Feshbach resonances. Each panel shows the measured phase shifts for a mean
collision energy of 656 and 798 nK \footnote{These mean collision energies are
weighted by $E_{\rm c}^{-1}$ to account for the energy dependence of
$(\pi^2/k^2)T_3T_4^*$ near $1\,\mu$K. For $|3(4), \pm 1>$, the
selection width is 13 nK and the mean energies are 644(616) and 771(746) nK.
The spread of collision energies for low and high energy are 428
and 461 nK for $m_F\neq \pm 1$ and 416(407) and 437(434) nK for $|3(4), \pm
1>$.}, which is changed by selecting a different detected velocity of the
scattered clock atoms. Results for target atoms with positive or negative $m_F$
are shown at magnetic fields with opposite signs, producing plots that are
continuous through $B=0$. The Feshbach resonances for 656 nK occur at lower
magnetic fields than those for 798 nK, and we observe slightly larger phase
variations through the resonances, as expected from the smaller spread of
collision energies. The error bars are the quadrature sum of the statistical
and systematic uncertainties, typically 30 mrad for 10 minutes of averaging at
points far from resonances. Through the resonances, where the scattering cross
section passes through zero, they may be as large as 100 mrad after 20 minutes
of averaging.

There are distinct similarities between the resonance positions and profiles
for target atoms $\vert 3, \pm|m_F| \rangle$ and $\vert 4, \mp(|m_F|+1)
\rangle$. In Figs.~\ref{fig:fig2}a)-g), we observe two clear resonant features.
For target atoms in $\vert 3, 3 \rangle$ and $\vert 4, -4 \rangle$ these
resonances are near 20 and 180 mG. For each of the other target states, one
resonance is near 50 mG and the other near $-80$ mG. We do not expect any
resonances for target atoms in $\vert 3, -3 \rangle$ and $\vert 4, 4 \rangle$
because conservation of angular momentum prohibits coupling to any closed
s-wave channels with halo states. While we observe only two resonant features
for each $\vert m_F \vert$, there are additional resonances that are not
resolved, because they overlap or are too narrow. For example, the results in
Ref.\ \cite{Gensemer12} indicate that there are two Feshbach resonances for
$\vert 3, 2 \rangle$ target atoms, while here we see only one. We show
experimentally that these resonances are in scattering channels with clock and
target atoms in different hyperfine levels, e.g.\ $\vert 4, 0 \rangle$ and
$\vert 3, 3 \rangle$, by measuring velocity-changing cross sections
\cite{Gibble95}.

The amplitude of the phase variation is different for each
resonant feature. We observe variations of nearly $\pi$ for some resonances,
but others produce variations as small as a few hundred mrad. The scattering
phase shift wraps through $\pi$ across an elastic Feshbach resonance, but even
in the elastic limit we will observe a smaller variation if the resonance is
narrower than our spread of collision energies. Inelastic loss may also reduce
the amplitude of the phase variations. Additional sharp phase changes may be
caused by the closing of inelastic scattering channels, but these are usually
smaller and do not wrap through $\pi$.

Fig.~\ref{fig:fig2} also shows the results of coupled-channel calculations
performed with the MOLSCAT quantum scattering package \cite{molscat:v14}, using
the interaction potentials of Berninger {\em et al.}\
\cite{Berninger:Cs2:2013}. The experimental observable is the Ramsey fringe in
Fig.~\ref{fig:fig1}b, which results from the interference of the scattered
atoms only, given by a quantity $J=\langle |f_{3,j}+f_{4,j}|^2 \rangle$
\cite{Kokkelmans_thesis}. Here the brackets denote an energy average, and
$f_{3,j}$ and $f_{4,j}$ indicate the scattering amplitudes for atoms in states
3 and 4, respectively, colliding with an atom in state $j$. The phase of the
fringe is shifted by the effect of the collisions, and is directly related to
the interference term in $J$. The phase shift can therefore be expressed as
$\Phi_j= \arg \langle T_{3,j} T_{4,j}^* \rangle$, where $T_{3,j}$ and $T_{4,j}$
are the diagonal $T$-matrix elements corresponding to the scattering amplitudes
$f_{3,j}$ and $f_{4,j}$.

The $T$-matrix elements may be written exactly in terms of complex
$k$-dependent scattering lengths $a$, $T=2ika/(1+ika)$ \cite{Hutson:res:2007}.
This gives
\begin{equation}
\Phi_j =
\arg\left\langle\left(\frac{2ika_{3,j}}{1+ika_{3,j}}\right)
\left(\frac{-2ika_{4,j}^*}{1-ika_{4,j}^*}\right)\right\rangle.
\label{eq:2nd}
\end{equation}
Writing $a=\alpha-i\beta$, $1+ika=1+k\beta+ik\alpha$ has a phase
$\arctan[k\alpha/(1+k\beta)]$. If the range of energies is narrow,
Eq.~(\ref{eq:2nd}) reduces to
\begin{eqnarray}
\Phi_j&\approx&
-\arctan\left(\frac{k\alpha_{3,j}}{1+k\beta_{3,j}}\right)
+\arctan\left(\frac{k\alpha_{4,j}}{1+k\beta_{4,j}}\right)
\nonumber\\
&+&\arg(a_{3,j})-\arg(a_{4,j}).
\label{eq:simplified}
\end{eqnarray}
When the scattering is purely elastic, $a$ is real, and Eq.\
\ref{eq:simplified} reduces to the difference between the scattering phase
shifts $\Phi_j=\delta_{4,j}-\delta_{3,j}$, with $\delta = -\arctan ka$. At zero collision energy,
$\delta_{4,j}-\delta_{3,j}$ vanishes but, in the presence of
inelasticity, the phases $\arg(a)$ contribute to $\Phi_j$ and persist to zero
energy. Note that our coupled-channel calculations evaluate the full expression
(\ref{eq:2nd}) for $\Phi_j$, including inelastic contributions.

The coupled-channel calculations are in overall excellent
agreement with the experimental results. The resonance positions and profiles
are well reproduced. Away from the Feshbach resonances, the background
phase-shift differences $\Phi_{4, m_F}$ depend weakly on collision energy and
agree quite well with the theoretical model. However, those for $\vert 3, m_F
\rangle$ target atoms show significant energy dependences and small but
statistically significant differences with the theoretical model.

To estimate the uncertainty in the predictions of the fitted potential, we have
repeated the fits of ref.\ \cite{Berninger:Cs2:2013} and determined
uncorrelated directions in the 6-parameter space. We have then found a
potential shifted in each of these directions by an amount that doubles the sum
of squares of residuals $\chi^2$ for the original data set of ref.\
\cite{Berninger:Cs2:2013}. For a locally linear fit, these correspond to
approximately $5\sigma$ uncertainties. We have repeated the coupled-channel
calculations of $\Phi_j$ for these 6 potentials. The differences from the
best-fit potential are small, and are shown for $\vert 3, \pm 3 \rangle$ in
Fig.~\ref{fig:fig2}h), together with corresponding differences for the
experimental results. For other targets, the differences between the shifted
potentials and the best-fit potential are even smaller. We conclude that the
remaining differences between experiment and theory are well outside the range
of the uncertainties from the interaction potential derived from the
experiments of Ref.\ \cite{Berninger:Cs2:2013}.

Fig.~\ref{fig:fig2}h) also shows the results obtained from coupled-channel
calculations using the previous best potential \cite{Chin04}, also plotted as
differences from the best-fit potential. For $\vert 3, \pm 3 \rangle$ and the
other target states. The potential from \cite{Berninger:Cs2:2013} gives
substantially better agreement through the resonances. The details of the bound
states that cause the low-field resonances are beyond the scope of this paper.
In essence, however, there is a group of pure triplet states bound by only 3.7
kHz at zero field that, as a function of magnetic field, are far
from parallel to the atomic thresholds below 0.1~G. Their crossings with the
thresholds cause the resonances we observe. At higher fields they mix with more
deeply bound states that possess some singlet character, and eventually become
almost parallel to the atomic thresholds at fields above 0.3~G.

To achieve the accuracy of these measurements, the experimental
sequence above avoids and accounts for several systematic errors. The largest
remaining systematic correction applied to the data in Fig. 2
comes from the interference between the scattered and unscattered waves.
This gives the usual loss of atom current in the forward
scattering direction, producing the dip in the distribution in
Fig.~\ref{fig:fig1}d) and contributing a different phase to the scattered
Ramsey fringe in Fig.~\ref{fig:fig1}b). We determine this contribution as a
function of the probed velocity: the phase shift of the interference current is
approximately zero and therefore, when $\Phi_j$ is far from 0, the correction
can be significant \cite{ABThesis}. For the background $\Phi_{3, m_F=(1,2,3)}$,
this correction is about (80,80,120) mrad for the low energy and (40,70,100)
mrad for high energy, increasing $\Phi_{3, m_F}$ (closer to the
theory) with a typical uncertainty of 25 mrad. The differences in
Fig.~\ref{fig:fig2}h for $\vert 3, \pm 3 \rangle$ are significantly larger than
this systematic uncertainty. Another significant systematic
arises because the scattered atoms experience a cold collision frequency shift
from the target atoms \cite{Gibble93,Papoular12}, in addition to the differential scattering phase shift.
Our sequence evaluates and corrects for this collision shift by measuring the
collision shift of the unscattered atoms (forward direction) due to the target
atoms \cite{ABThesis,Note1}. The correction is typically $-40(0)\pm 3$ mrad for $\Phi_{3(4), m_F}$. We also apply a small correction due to inelastic
spin-changing collisions populating other $\vert F,m_F\rangle$ target states
\cite{ABThesis,Note1}.

In summary, we precisely measure quantum scattering phase shifts spanning a
series of Feshbach resonances and compare them to a state-of-the-art
theoretical model. These results provide a stringent confirmation of the cesium
interaction potentials of ref.\ \cite{Berninger:Cs2:2013}, but small,
statistically significant differences remain unexplained. We have considered
the uncertainties in the theoretical predictions due to statistical
uncertainties in the fitted interaction potentials and shown them to be very
small. The theory shows that inelastic processes make important contributions
to the observable quantities that persist even in the limit of zero collision
energy. With this experimental technique, we can currently determine
differential scattering phase shifts with mrad precision in less than a day of
averaging. Further work using these and further improved
interaction potentials may probe how this technique can best set stringent
limits on the time variation of fundamental constants, such as the
electron-proton mass ratio, by observing the constancy of the scattering phase
shifts near narrow Feshbach resonances \cite{Chin06, Borschevsky11}.

\begin{acknowledgments}
We are grateful to S. Gensemer for contributions during the initial stages of
this experiment, to C. R. Le Sueur for work to modify MOLSCAT to handle
asymptotically degenerate states, and to E. Tiesinga for stimulating
discussions. We acknowledge financial support from the NSF, NASA, Pennsylvania
State University and the UK Engineering and Physical Sciences Research Council
under grant numbers EP/I012044/1, EP/N007085/1, and EP/P01058X/1. This work is
part of the Vici research programme with project number 680-47-623, which is
financed by the Netherlands Organisation for Scientific Research (NWO).

\end{acknowledgments}

\bibliography{Biblio-Database2,jmh-bib}

\pagebreak

\begin{center}
\textbf{\large Supplemental Material}
\end{center}
\setcounter{equation}{0}
\setcounter{figure}{0}
\setcounter{table}{0}
\setcounter{page}{1}
\makeatletter
\renewcommand{\theequation}{S\arabic{equation}}
\renewcommand{\thefigure}{S\arabic{figure}}

\section{Experimental Apparatus}

\begin{figure}
  \centerline{\includegraphics[width=\columnwidth]{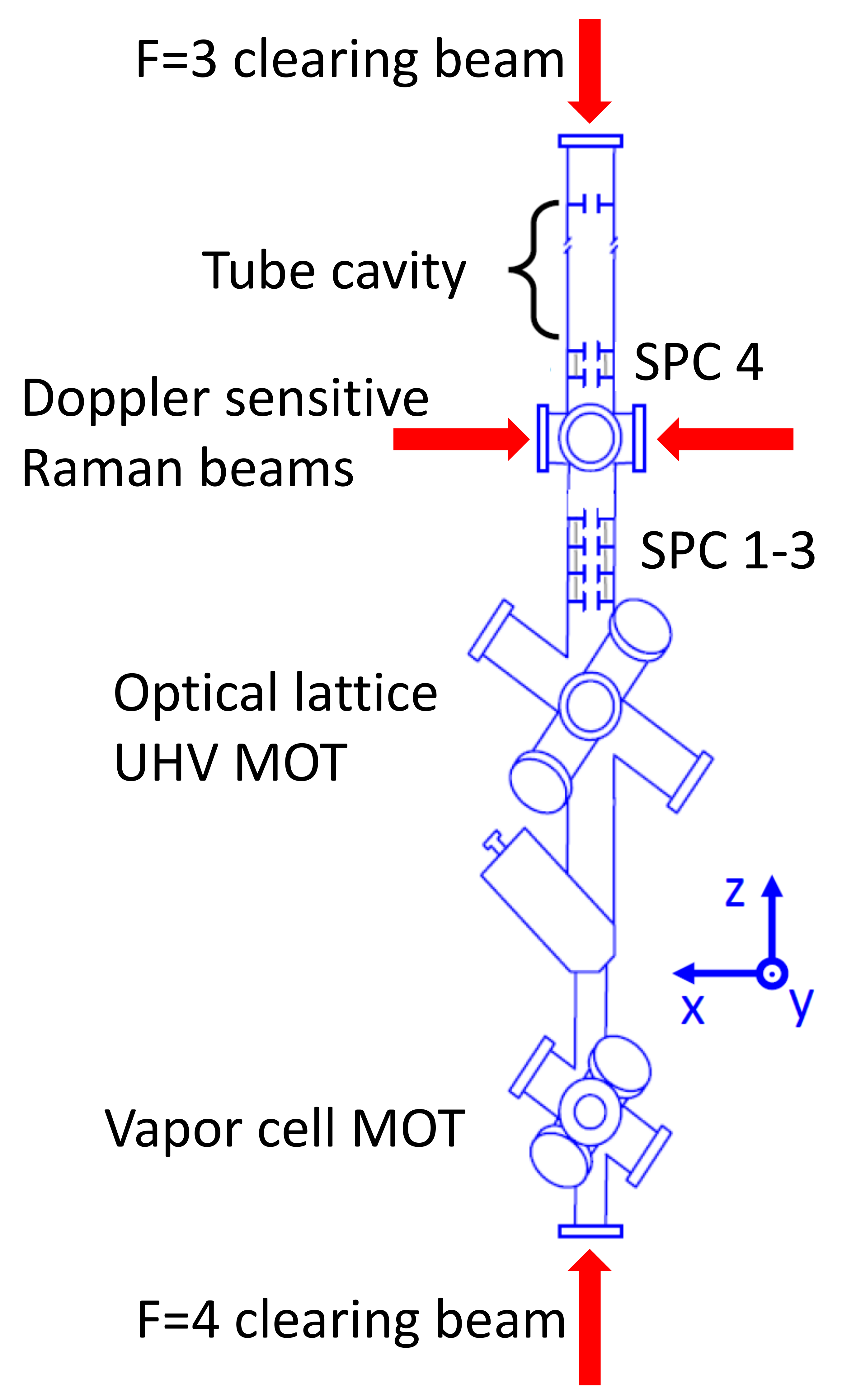}}
\caption[Probe Velocities for Interference Current Check.]{\label{fig:SMAppar}
(color online). Cesium atomic fountain clock schematic.}
\end{figure}

Our cesium atomic clock is depicted in Fig.~\ref{fig:SMAppar}. It has
previously been described in \cite{Hart07,Gensemer12}. Atoms are collected in
the vapor cell magneto-optical trap (MOT) and multiply launched into the
ultra-high vacuum (UHV) MOT. They are then launched and loaded into a
moving-frame optical lattice to optically pump and cool the atoms to 400 nK.
State preparation cavities (SPC) resonant at 9.2 GHz drive microwave
transitions to prepare the atoms in specific $\vert F, m_F \rangle$ states. The
Raman beams counter-propagate and select the clock atoms as they travel
upwards, followed by probing the atoms as they return from the clock cavity.
The clearing beams push atoms in the $F=3$ or $F=4$ hyperfine states to remove
them from their fountain trajectory. In this experiment, the tube cavity serves
as the clock cavity, exciting the atoms at its lowest field antinode. This
fountain has no magnetic shielding for static magnetic fields. A long aluminum
cylinder, with a wall thickness of 0.5 inches, surrounds SPC 4 and the tube
cavity, to shield a small ambient 60 Hz magnetic field and its higher
harmonics.

\section{Experimental Sequence for $\vert 3, \pm1 \rangle$ targets}

For $\vert 3, \pm 1 \rangle$ targets, our sequence draws the target and clock
atoms from $\vert 3,3 \rangle$. Microwave pulses first transfer atoms launched
in $\vert 3,3 \rangle$ to $\vert 4,0 \rangle$, followed by a Raman selection
that transfers a portion of the atoms to $\vert 3, \mp 1 \rangle$. A microwave
pulse then transfers the unselected atoms in $\vert 4,0 \rangle$ to $\vert 3,
\pm1 \rangle$. We then clear F=4 and transfer the selected atoms back to $\vert
4,0 \rangle$. With the target atoms in $\vert 3, \pm 1 \rangle$ and the clock
atoms in $\vert 4,0 \rangle$, the sequence follows that for other $\vert 3, m_F
\rangle$ targets.

\section{Systematic errors}

\subsection{Frequency Shift due to Ultracold Collisions}

We ascertain the cold collision shift using additional experimental sequences
to measure the Ramsey fringes of the unscattered atoms with target atoms
present and with the target atoms cleared just before the first Ramsey pulse.
However, because the scattered and unscattered atoms have slightly different
fountain trajectories, this measured correction is not exact. We therefore also
measure $\Phi$ as a function of target atom density and extrapolate the
measured phase shifts to zero density to confirm independently that the error
of this correction is safely less than the statistical error bars
\cite{ABThesis}. Because these collision shifts result from an interference
between the scattered and incident atom waves, their values depend on the
scattering phase shifts and we therefore observe significant variations of the
collision shifts through Feshbach resonances. Ref.\ \cite{Papoular12} reported
related measurements of the ratios of the frequency shift for clock state and
$F=3$ target atoms between $-80$ and 100 mG.

\subsection{Corrections due to Inelastic Collisions}

\begin{figure}
  \centerline{\includegraphics[width=\columnwidth]{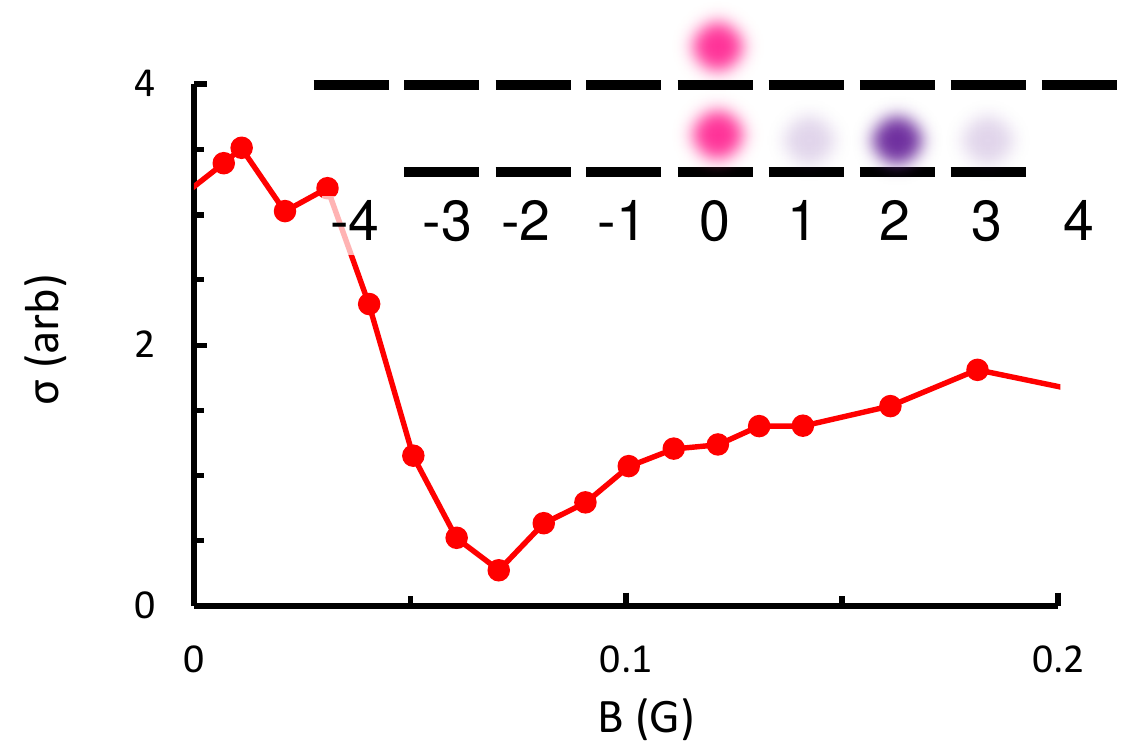}}
\caption[Probe Velocities for Interference Current Check.]{\label{fig:Figure3}
(color online). Inelastic collisions between target atoms in $\vert 3,2
\rangle$ populate the $\vert 3,1 \rangle$ and $\vert 3,3 \rangle$ states. When
the scattering cross section for collisions between the intended clock and
target state goes to zero, the fractional contribution to the scattered Ramsey
fringes from clock atoms colliding with the inelastically populated states can
be significant.}
\end{figure}

We also correct our measurements of $\Phi$ for a systematic error due to
inelastic spin-changing collisions that populate other $\vert F,m_F \rangle$
target states. For example, inelastic collisions between target atoms in $\vert
3, 2 \rangle$ populate $\vert 3,1 \rangle$ and $\vert 3,3 \rangle$ (see Fig.\
\ref{fig:Figure3}). Inelastic collisions that conserve $F$ produce a systematic
error because $\Phi_{3,1}$ and $\Phi_{3,3}$ are different from $\Phi_{3,2}$.
Our sequence minimizes this error by transferring the target atoms to the
target state just before the first Ramsey pulse and clears atoms that scattered
inelastically up to that point in the fountain trajectory. Because the number
of inelastic collisions is small, this error is only significant when the
scattering cross section of the intended target and clock states approaches
zero through a Feshbach resonance, as shown in Fig.\ \ref{fig:Figure3} for
$\vert 3, 2 \rangle$ target atoms. Similar inelastic collisions occur between
clock and target atoms, but because the clock atom density is $\approx20$ times
smaller than the target cloud density, populations in unintended states from
inelastic collisions involving clock atoms are negligible. Between 50 and 100
mG, clock atoms colliding with $\vert 3, 1 \rangle$ and $\vert 3, 3 \rangle$
contribute significantly to the phase of the scattered Ramsey fringe. The
results in Fig.\ \ref{fig:fig2} include this correction, as large as 75 mrad
and comparable to the statistical error bars of about 100 mrad \cite{ABThesis}.

\subsection{Magnetic field stability and uniformity}

To reduce background fluctuations in the magnetic field, we actively stabilize
all three components of the vector field using a flux gate magnetometer and a
digital servo. The residual field noise is $\pm 50~\mu$G, which
produces noise due to the quadratic Zeeman shift at a bias field of 0.4~G that
is below our background noise. We also minimize gradients of the magnetic
field in the flight region, which can give a systematic error since the atoms
that produce the scattered and unscattered Ramsey fringes in Fig.\
\ref{fig:fig1}b follow different trajectories and therefore experience
different quadratic Zeeman shifts. The field is in the vertical ($z$) direction
so we minimize the gradients $\frac{dB_z}{dz}, \frac{dB_z}{dy},$ and
$\frac{dB_z}{dx}$. To reduce $\frac{dB_z}{dz}$, we measure the $\vert 3, 3
\rangle$ to $\vert 4, 3 \rangle$ transition frequency at various heights above
the clock cavity and apply correction currents to coils surrounding the
apparatus. To measure and null $\frac{dB_z}{dy}$, we drive the same
magnetically sensitive transition but mask opposite halves of the detection
beam on successive shots, probing the field in either the positive or negative
$y$ half of the fountain. To measure and minimize $\frac{dB_z}{dx}$, we use
photon recoils from a pair of Raman beams to push the cloud in opposite
directions before driving the magnetically sensitive transition. We then adjust
the field so that the atoms experience the same magnetic field whether they are
pushed in the positive or negative $x$ direction. We subsequently observe
Ramsey fringes for recoils in opposite directions with phase differences less
than 10 mrad at 400 mG \cite{ABThesis}.

\end{document}